\documentclass[letter]{aa}
\usepackage{graphicx}
\usepackage{dirtytalk}
\usepackage{currvita}
\usepackage[T1]{fontenc}
\usepackage{txfonts}
\usepackage{hyperref}
\usepackage{xcolor}
\hypersetup{
colorlinks,
linkcolor={red!80!black},
citecolor={blue!80!black},
urlcolor={blue!80!black}
}

\newenvironment{myfont}{\fontfamily{ppl}\selectfont}{\par}
\DeclareTextFontCommand{\textmyfont}{\myfont}

\newcommand{\code}[1]{\texttt{#1}}

\def\nifs{\iso{56}Ni}

\def\cm3{cm$^{-3}$}

\def\msun{$M_{\odot}$}

\def\one{\ts {\,\sc i}}
\def\two{\ts {\,\sc ii}}
\def\three{\ts {\,\sc iii}}

\def\beq{\begin{equation}}
\def\eeq{\end{equation}}

\def\lesssim{\mathrel{\hbox{\rlap{\hbox{\lower4pt\hbox{$\sim$}}}\hbox{$<$}}}}
\def\gtrsim{\mathrel{\hbox{\rlap{\hbox{\lower4pt\hbox{$\sim$}}}\hbox{$>$}}}}

\def\one{{\,\sc i}}
\def\two{{\,\sc ii}}
\def\three{{\,\sc iii}}

\def\cmfgen{{\code{CMFGEN}}}

\def\wiserep{{\code{WISEREP}}}

\def\lsst{LSST}

\def\niidoub{[N\two]\,$\lambda\lambda$\,$6548,\,6583$}

\def\caiidoub{[Ca\two]\,$\lambda\lambda$\,$7291,\,7323$}
\def\caiitrip{Ca\two\,$\lambda\lambda\,8498-8662$}

\def\oidoub{[O\one]\,$\lambda\lambda$\,$6300,\,6364$}

\def\oitrip{O\one\,$\lambda\lambda$\,$7771-7775$}

\def\mgi{Mg\one]\,$\lambda\,4571$}
\def\nad{Na\one\,$\lambda\lambda\,5896,5890$}

\newcommand{\iso}[2]{\ensuremath{^{#1}\rm{#2}}}

\begin{document}

   \title{Using LSST late-time photometry to constrain  Type Ibc supernovae and their
   progenitors}

  \titlerunning{SN Ibc properties inferred from LSST photometry}

\author{
Luc Dessart\inst{\ref{inst1}}
\and
 Jose~L. Prieto\inst{\ref{inst2},\ref{inst3}}
\and
 D. John Hillier\inst{\ref{inst4}}
\and
Hanindyo Kuncarayakti\inst{\ref{inst5},\ref{inst6}}
\and
Emilio~D. Hueichapan\inst{\ref{inst2}}
 }

\institute{
    Institut d'Astrophysique de Paris, CNRS-Sorbonne Universit\'e, 98 bis boulevard Arago, F-75014 Paris, France.\label{inst1}
\and
N\'ucleo de Astronom\'ia de la Facultad de Ingenier\'ia y Ciencias, Universidad Diego Portales,
Av. Ej\'ercito 441 Santiago, Chile.\label{inst2}
\and
Millennium Institute of Astrophysics, Santiago, Chile.\label{inst3}
\and
 Department of Physics and Astronomy \& Pittsburgh Particle Physics, Astrophysics, and Cosmology Center (PITT PACC),  \hfill \\
 University of Pittsburgh, 3941 O'Hara Street, Pittsburgh, PA 15260, USA.\label{inst4}
\and
    Tuorla Observatory, Department of Physics and Astronomy, FI-20014 University of Turku, Finland.\label{inst5}
 \and
     Finnish Centre for Astronomy with ESO (FINCA), FI-20014 University of Turku, Finland.\label{inst6}
}

   \date{}

  \abstract{
Over its lifespan, the Vera C. Rubin Observatory Legacy Survey of Space and Time (LSST) will monitor millions of supernovae (SNe) from explosion to oblivion, yielding an unprecedented $ugrizy$ photometric dataset on their late-time evolution. Here, we show that the photometric evolution of Type Ibc SNe can be used to constrain numerous properties of their ejecta, without the need for expensive spectroscopic observations. Using radiative-transfer simulations for explosions of He-star progenitors of different initial masses, we show that the $g$-band filter follows primarily the strength of the Fe\two\ emission, the $r$-band \oidoub\ and \niidoub, the $i$-band \caiidoub, and the $z$-band the \caiitrip\ triplet, and hence provides information on nucleosynthetic yields. Information on weaker lines, which may be used, for example, to constrain clumping,  is absent. However, this deficiency may eventually be cured by improving the physical realism of radiative-transfer simulations through a closer connection to physically consistent 3D explosion models, and by the judicial selection of a much smaller set of spectral observations. Degeneracies inherent to the SN radiation will affect the interpretation of photometric measures, but line fluxes from nebular-phase spectra are similarly compromised. Importantly, our ``family'' of Type Ibc SN models follows a distinct trajectory in color-color magnitude diagrams as the ejecta evolve from 100 to 450\,d, allowing one to disentangle different progenitors or explosions.  This photometric procedure provides a promising approach to study statistical samples of SNe Ibc and to confront them to ever improving progenitor and explosion models, to capture the onset of late-time interaction with circumstellar material, or to identify events currently unknown.
}


   \maketitle

\section{Introduction}

The supernova (SN) community is approaching an important turning point. In the last few decades, the discovery of massive star explosions has grown from a few per year to many each night (see, e.g., \citealt{sullivan_snrate_13}), thanks to numerous all-sky surveys such as PTF \citep{ptf}, Pan-STARRS \citep{panstarrs}, ASAS-SN \citep{shappee_assasn_14}, or ATLAS \citep{atlas_18}. In parallel, the modeling of SN radiation has expanded considerably, with the development of numerous tools to model  light curves and/or spectra. However, much of this modeling has focused on only a few well-observed nearby events. Because of obvious observational challenges, these studies are also biased in favor of the high brightness, photospheric phase rather than the low-brightness, ever dimming, nebular phase.

With the Vera C. Rubin Observatory Legacy Survey of Space and Time (LSST; \citealt{lsst_09}), the SN discovery rate will be further enhanced.  To maximize the use of this treasure-trove of data there is an urgent need to develop photometric techniques that can constrain SN properties, and be used to  provide constraints on stellar and galactic evolution, as currently attempted with spectroscopy. Photometry is less time consuming than spectroscopy, can be obtained for fainter objects, and is gathered in a systematic manner with modern, untargeted sky surveys.   Further, the very faint magnitude detection limit allows a monitoring of SNe out to very late times. A continuous photometric light curve of SNe can thus be built from maximum light until late times, yielding  information competitive with that obtained using spectra that  are much harder to obtain, and which have a much poorer temporal coverage.

The \lsst\ will yield an unprecedented number of SNe suitable for statistical analyses, with estimates of $\sim$\,3.3 million Type II SNe and 580 thousand Type Ibc SNe for a ten year survey. Within a redshift of 0.07, the total would be about 25 thousand Type II SNe and about eight thousand Type Ibc SNe.  About 4400/4300/3000 of such SNe Ibc will have \lsst\ photometric detections in $g$/$gr$/$gri$ at 100 days after maximum light. At 200\,d, these numbers change to 2300/2000/1700 and at 300\,d, they are 400/380/380. These estimates come from the Photometric LSST Astronomical Time Series Classification Challenge (PLAsTiCC; \citealt{plasticc_lsst_19,plasticc_lsst_20}) and are based at nebular times on the theoretical light curves of Dessart et al. (in preparation)\footnote{ZTF obtains $g$ and $r$ band photometry down to 20--20.5\,mag and therefore captures a subset of what LSST will provide in terms of depth and spectral coverage. Similarly, Pan-STARRS obtains $grizy$ photometry down to $\sim$\,21\,mag, and thus does not go as deep as the \lsst.}. In the absence of dust (and molecule emission) the coverage from $\sim$\,3600 ($u$ band) to $\sim$\,9700\,\AA\ ($y$ band) captures typically 70\% of the SN light at nebular times, and thus permits the determination of the bolometric luminosity, although reddening and distance (redshift) may not be trivial to estimate accurately. Correction for host contamination is also easier and more robustly done with difference imaging and photometry, as compared to typical slit spectroscopy.

During the photospheric phase, the bolometric luminosity, the brightness in various photometric filters, and the colors, provide constraints on the progenitor radius and the global properties of the ejecta such as its mass and kinetic energy. However they provide only limited information on the composition with the exception of \nifs\ (i.e., in ejecta where its decay provides the main energy source for the SN luminosity). In contrast, at nebular times, the ejecta is relatively optically thin, and the SN radiation is primarily seen as line emission. This is particularly evident in Type Ib (and IIb) and Ic SNe at about a year post explosion when most of the optical flux is contained in \oidoub\ and \caiidoub.

 Omitting the line profile morphology, the photometry yields very similar information to spectroscopy at late times -- the relative strength of important emission lines may be inferred from a well chosen optical color. For example, the  $r-i$ color contains essentially the same information as the flux ratio of the \oidoub\ and \caiidoub\ lines (modulo the contribution of the \niidoub\ line in the lower mass models; Fig.~\ref{fig_oi_nii_over_caii}). With radiative transfer modeling one can potentially constrain the mass of the elements associated with this line emission \citep{fransson_chevalier_89}. Although this has been done so far exclusively using the detailed information in spectra, the \lsst\ motivates the use of photometric measures alone to deliver similar constraints on the ejecta nucleosynthesis. With the large number of SNe that will be monitored for years by the \lsst, such a photometric approach may be applied to a statistical sample of SNe, constraining some properties of the ejecta and progenitor, as well as connecting these to the host populations that will also be resolved in finer detail than ever before (by the \lsst\ itself, as well as with the next generation large telescopes; see, e.g., \citealt{pisco_18}, \citealt{kuncarayakti_env_18}).

In the next section, we illustrate how photometry, and in particular colors, can be used to assess the relative strength of line emission in SNe Ibc at nebular times, allowing one to identify different types of ejecta composition based on photometry alone. In Section~\ref{sect_disc}, we discuss the limitation of this method and possible workarounds for these. In Section~\ref{sect_conc}, we present our conclusions.

\begin{figure}
\centering
\includegraphics[width=\hsize]{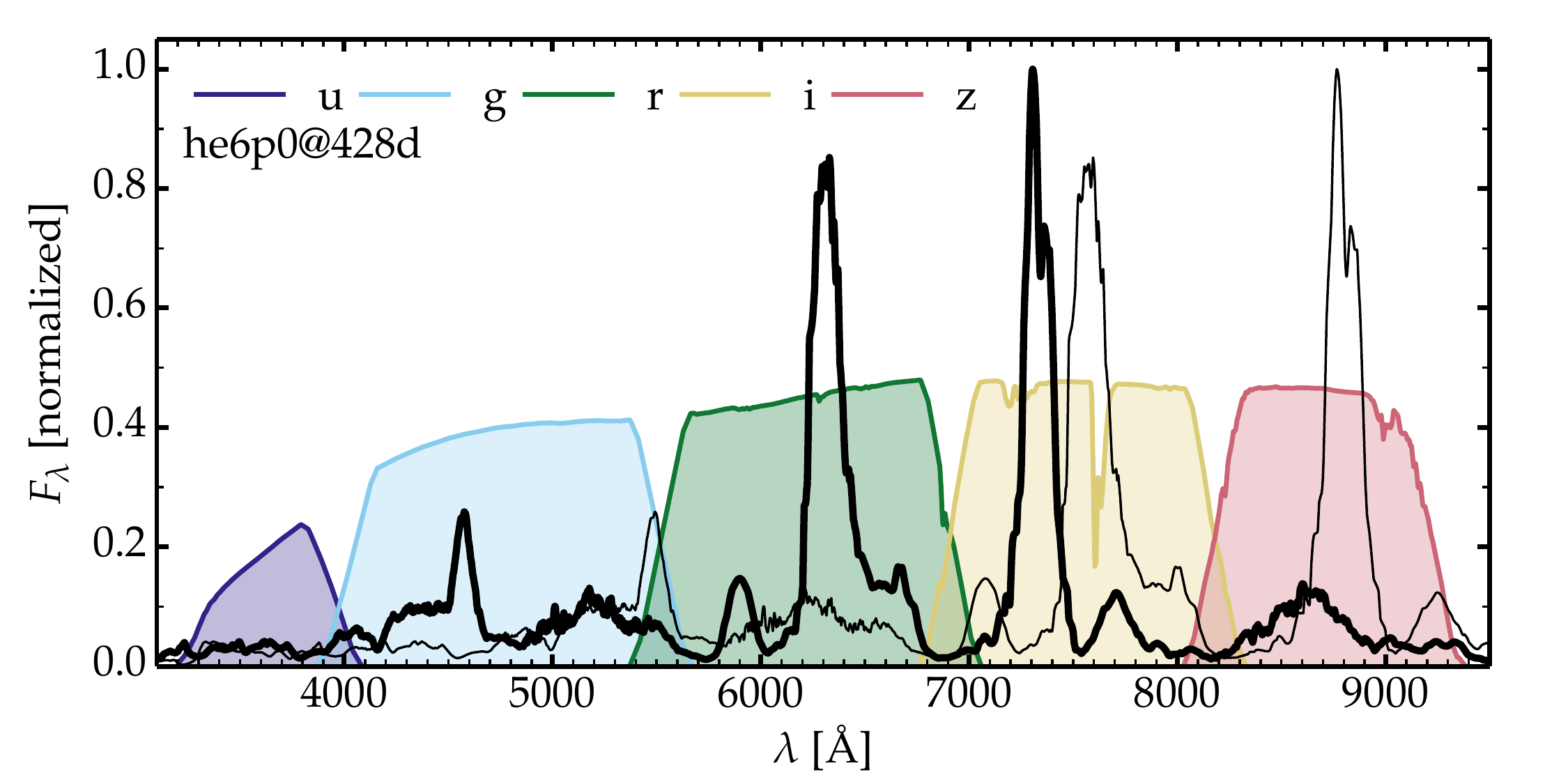}
\caption{Spectrum for  model he6p0 at 428\,d after explosion (thick black line; Dessart et al., in preparation) together with the normalized LSST $ugriz$ filter transmission functions (the $y$ filter is not included in that figure). The thin black line is the model spectrum for  a redshift of 0.2. For redshifts of $\sim$\,0.1 the photometric measures would be hard to use since \oidoub\ and \caiidoub\ fall between filters.
\label{fig_filters}
}
\end{figure}

\begin{table}
    \caption{Ejecta properties of our set of Type Ibc models from D21.
\label{tab_prog}
    }
    \begin{center}
      \begin{footnotesize}
  \begin{tabular}{l@{\hspace{2mm}}|c@{\hspace{2mm}}c@{\hspace{2mm}}c@{\hspace{2mm}}
      c@{\hspace{2mm}}c@{\hspace{2mm}}c@{\hspace{2mm}}c@{\hspace{2mm}}
    }
    \hline
Model  &  $M_{\rm ZAMS}$  & $M_{\rm preSN}$  & $M_{\rm ej}$  &     $E_{\rm kin}$  &  \iso{4}He  & \iso{16}O &  \iso{56}Ni$_{t=0}$  \\
       &     [\msun] & [\msun] & [\msun]                 &        [foe]     &    [\msun] & [\msun] & [\msun]   \\
\hline
   he2p6  & 13.85  & 2.15 &   0.79  &     0.13   &    0.71  &   2.28(-2)     &   1.22(-2)     \\
   he2p9  & 14.82  & 2.37 &   0.93  &     0.37   &    0.77  &   5.03(-2)     &   2.32(-2)     \\
   he3p3  & 16.07  & 2.67 &   1.20  &     0.55   &    0.84  &   1.51(-1)     &   4.00(-2)     \\
   he3p5  & 16.67  & 2.81 &   1.27  &     0.41   &    0.87  &   1.72(-1)     &   2.92(-2)     \\
   he4p0  & 18.11  & 3.16 &   1.62  &     0.63   &    0.92  &   3.10(-1)     &   4.45(-2)     \\
   he4p5  & 19.50  & 3.49 &   1.89  &     1.17   &    0.95  &   4.19(-1)     &   8.59(-2)     \\
   he5p0  & 20.82  & 3.81 &   2.21  &     1.51   &    0.97  &   5.92(-1)     &   9.77(-2)     \\
   he6p0  & 23.33  & 4.44 &   2.82  &     1.10   &    0.95  &   9.74(-1)     &   7.04(-2)     \\
   he7p0  & 25.68  & 5.04 &   3.33  &     1.38   &    0.90  &   1.29(0)      &   1.02(-1)     \\
   he8p0  & 27.91  & 5.63 &   3.95  &     0.71   &    0.84  &   1.71(0)      &   5.46(-2)     \\
   he12p0 & 35.74  & 7.24 &   5.32  &     0.81   &    0.23  &   3.03(0)      &   7.90(-2)     \\
 \hline
  \end{tabular}
  \end{footnotesize}
\end{center}
    {\bf Notes:} The table columns correspond to the ZAMS mass, the preSN mass, the ejecta mass, the ejecta kinetic energy (1 foe $\equiv
    10^{51}$\,erg), and the cumulative yields of \iso{4}He, \iso{16}O, and \nifs\ prior to decay (see discussion in D21). Numbers in parentheses represent powers of ten.
\end{table}

\section{Colors as a proxy for line-flux ratios}
\label{sect_res}

Figure~\ref{fig_filters} shows the $ugriz$ \lsst\ filters together with a nebular spectrum for a SN ejecta arising from the explosion of a star that started on the He zero age main sequence with a mass of 6\,\msun\ (\citealt{dessart_snibc_21}; hereafter D21; this model would most likely have been classified as a SN Ib; \citealt{dessart_snibc_20}). This spectrum is dominated by lines of  moderate to large strength such as \mgi, \nad, \oidoub, \niidoub, \caiidoub, \oitrip, and \caiitrip, together with many weak Fe\two\ lines contributing a moderate but extended background emission between 4000 and 5500\,\AA. In some models characterized by low ionization, Fe recombines, and the Fe\two\ emission in the blue disappears in favor of Fe\one\ emission further to the red (D21).

The emission from low-redshift Type Ibc SNe ($z \lessapprox 0.07$) is captured in a ``convenient way'' by the \lsst\ photometric filters since the landmarks of their nebular spectra fall in distinct filters. The strong Fe\two\ emission, which arises from the He-rich shell at all times or from the O-rich shell at early times, falls in the $g$ band. \oidoub\ falls in the $r$ band, together with weaker contributions from Fe\two\ lines early on and by \niidoub\ later on in progenitors that retained their  He/N shell all the way to core collapse. \caiidoub\ falls in the $i$ band and dominates over other contributions from Fe\two\ or Ni\two, while \caiitrip\ falls in the $z$ band. Finally, we mention the $u$ band which will provide insights into the possible interaction of the ejecta with circumstellar material.  In Type II SNe, H$\alpha$ is always strong and falls in the same filter as \oidoub\ so the [O\one] line flux contribution to the $r$-band magnitude cannot easily be made in H-rich ejecta with such photometric filters. Type II SNe are thus not included in the following discussion.

\begin{figure}
\centering
\includegraphics[width=\hsize]{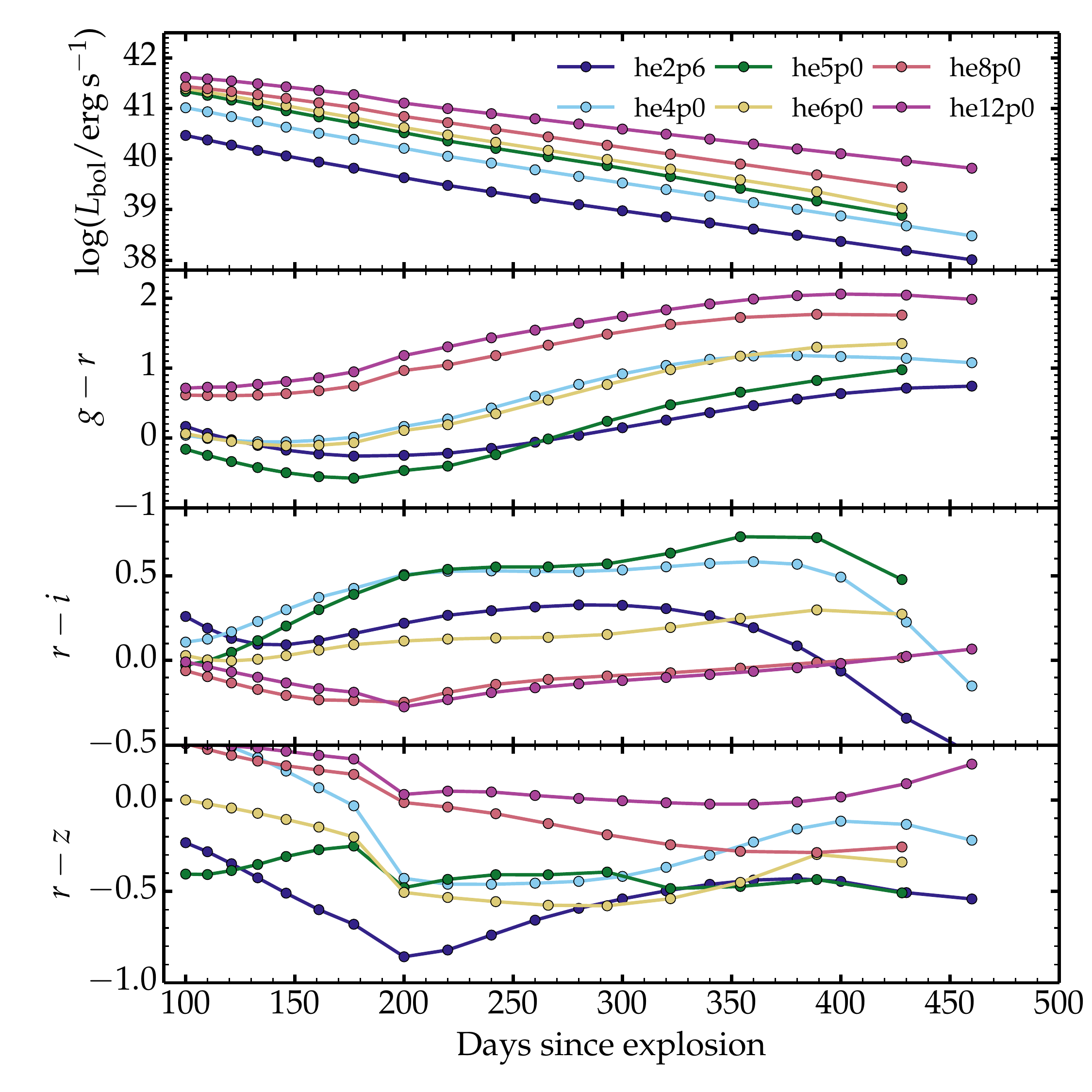}
\caption{Photometric properties of a sample of He-star explosion models from 100 to about 450\,d after explosion. The bolometric light curve is shown at top, followed by the color curves $g-r$, $r-i$, and $r-z$. The original models are presented in D21, but their full time evolution is discussed in Dessart et al. (in preparation).
\label{fig_phot}
}
\end{figure}

Of interest here is whether one may constrain the progenitors and  yields of Type Ibc SNe based on photometry alone. In other words, can photometry alone distinguish robustly progenitors that died with different abundances of O, He, or N. Practically, this requires that differences in line fluxes seen in explosion models of different age, composition, energetics, correspond to clear differences in photometric properties. Degeneracies that lead to similar line fluxes or line flux ratios from distinct models will impact both photometric and spectroscopic measures --- this issue concerns the physics of SN ejecta and the degeneracy of the SN radiation itself. Extensive time coverage that is available with photometry may help to break some of these degeneracies.

To test the usefulness of photometry, Dessart et al. (in preparation) employed the He-star explosion simulations of D21 and evolved these from an earlier time of 100\,d (rather than 200\,d) to about 450\,d after explosion with the nonlocal thermodynamic equilibrium time-dependent radiative transfer code \cmfgen\ \citep{HD12}. The sample of models includes initial He-star masses of 2.6, 2.9, 3.3, 3.5, 4.0, 4.5, 5.0, 6.0, 7.0, 8.0, and 12.0\,\msun\ (the models are named he2p6 etc.; see Table~\ref{tab_prog}, \citealt{woosley_he_19}, and \citealt{ertl_ibc_20}). An extensive description of the resulting spectra and ejecta properties is deferred to a forthcoming study (Dessart et al., in preparation)  but since it can help understand what drives the photometric changes, we show the spectral evolution for a low-mass and a high-mass He-star model in the appendix (Fig.~\ref{fig_spec}). Here, we focus on their photometric properties and we selected only a subset of representative models, namely he2p6, he4p0, he5p0, he6p0, he8p0, and he12p0. The D21 models with clumping throughout the ejecta or only in the O-shell are excluded for conciseness.

\begin{figure}
\centering
\includegraphics[width=\hsize]{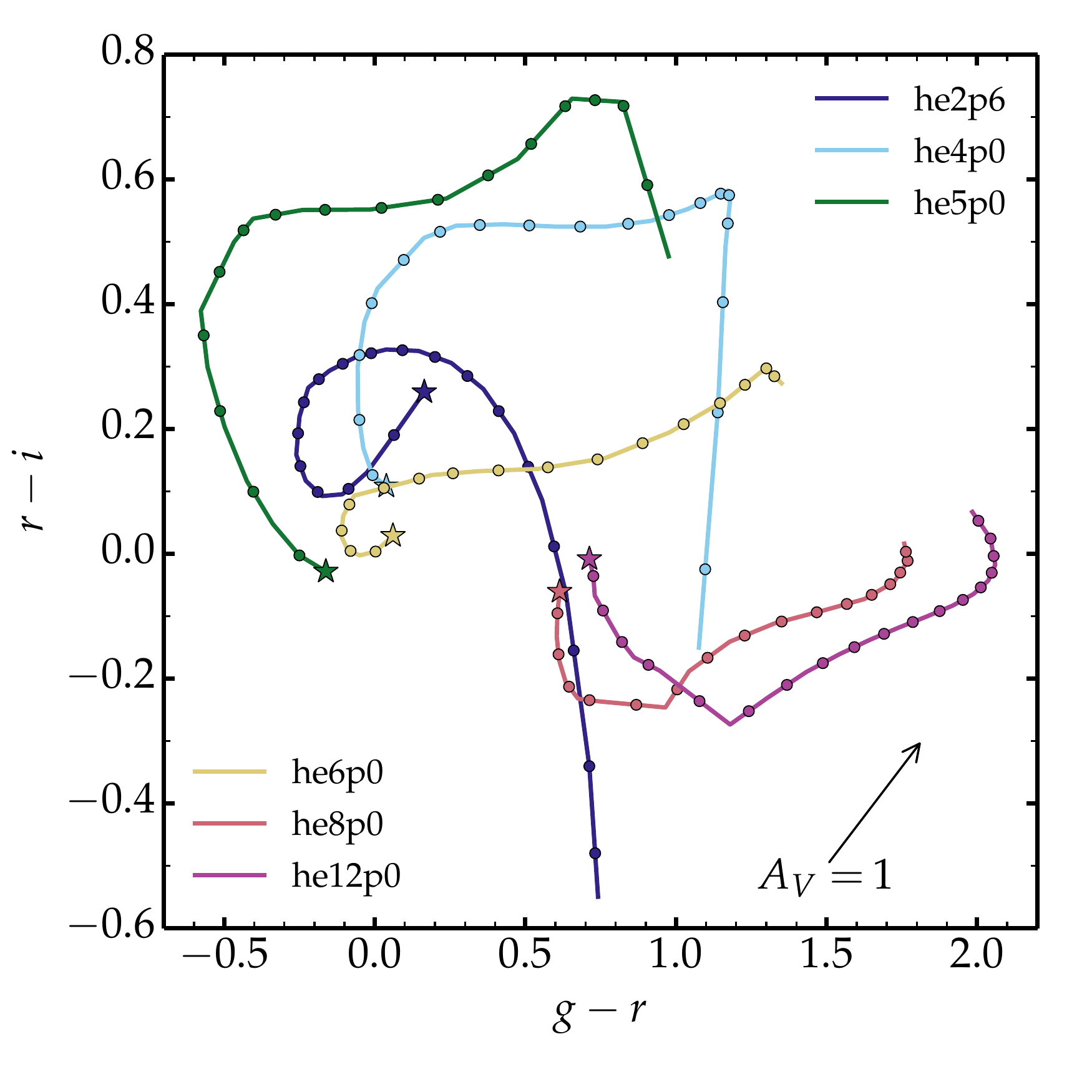}
\vspace{-1cm}
\caption{LSST color-color magnitude diagram for the Type Ibc simulations of Dessart et al. (in preparation) from 100 to about 450\,d after explosion. We show the color curves $r-i$ versus $g-r$ for our He-star explosion models from 100\,d (indicated by a star symbol) until the end of the simulation at around 450\,d. Dots are equally spaced in time every 20\,d. The arrow corresponds to the color shift caused by $A_V=$\,1\,mag.
\label{fig_color}
}
\end{figure}

Some photometric properties of our models are shown in Fig.~\ref{fig_phot}. Because the kinetic energy tends to increase with ejecta mass, the fading rate is similar between models. The decay power varies by about a factor of ten (the initial \nifs\ mass covers the range 0.01 up to 0.1\,\msun), while in all models the fraction of $\gamma$ rays that escape increases by a factor of about five between 100 and 450\,d. Hence, the bolometric light curves appear ordered in mass and do not cross as time passes. When considering photometry, and especially observations, the magnitude in a given band will depend on, for example, the \nifs\ mass, and the distance, reddening, and redshift of the SN. In contrast, working with colors directly takes out the issue with distance, and takes out some of the influence of variations in \nifs\ mass. So, color is a more direct way to distinguish the models.

The bottom three panels of Fig.~\ref{fig_phot} illustrate the color evolution in $g-r$, $r-i$, and $r-z$. The $g$-band brightness reflects primarily the strength of Fe\two\ emission, with a growing contribution from \mgi\ at late times. D21 found that this Fe\two\ emission is relatively strong in the lower mass He-star models, since Fe\two\ is the primary coolant for the He-rich shell (which represents 90\% of the ejecta mass in model he2p6) or for the O-rich shell when O is partially ionized (model he5p0 here). The $r$ band contains \oidoub, \niidoub, as well as Fe\two\ line emission. The latter is strong around 100\,d and weakens in time. In contrast, \niidoub\ (\oidoub) strengthens continuously from 100 to 450\,d in models with a massive He-rich (O-rich) shell. Models of greater mass, or of lower expansion rate, tend to be more recombined and thus lie at higher $g-r$ values (redder color).

The $i$ band is mostly sensitive to \caiidoub, which forms in the Fe/He and Si/S shells of the ejecta (\citealt{jerkstrand_15_iib}; D21).\footnote{In contrast to \citet{li_mccray_ca2_93}, \citet{D21_sn2p_neb} find that \caiidoub\ also forms in the Fe/He and Si/S shells in Type II SN ejecta.} The slowly growing $r-i$ color indicates a progressive though modest strengthening of \caiidoub\ relative to \oidoub. In models with a He-rich shell,  \niidoub\ becomes very strong at late times and drives $r-i$ to negative values. This effect is very strong in model he2p6, so much so that one could mistake \niidoub\ with H$\alpha$, as observed in SN\,1993J \citep{matheson_93j_00b}. Finally, the $z$ band is a good tracer of \caiitrip. In lower mass He-star models, this triplet is weak because Ca is overionized through most of the ejecta (i.e, the He-rich shell). In higher mass models, \caiitrip\ forms in the O-rich shell and is strong at earlier times when the ejecta is not too optically thin. In all cases, the \caiidoub\ is stronger at late times (for additional discussion and details, see D21 and Dessart et al., in preparation).

A more direct way to distinguish the models shown in Fig.~\ref{fig_phot} is to consider the color-color magnitude diagram $r-i$ versus $g-r$ (Fig.~\ref{fig_color}). Not only do the models lie in different parts of the diagram, but the trajectories they follow as they evolve are distinct. Lower mass models lie at small (negative) $g-r$ values, only moving to the right at late times, preferentially at small (negative) $r-i$ values because of the strong flux in the $r$ band due to \oidoub\ or \niidoub. Higher mass models, with a greater O yield, are systematically redder and lie at higher $g-r$ values, with a small $r-i$ value that reflects the dominance of  \oidoub\ over the optical spectrum. These photometric properties are based on a subset of the He-star explosion models of D21, which have their own idiosyncrasies. The present set of trajectories is primarily illustrative but it shows that this color-color magnitude diagram captures the basic spectral families drawn out by these models -- the photometry conveys rich information similar to that contained in spectra.

\begin{figure}
\centering
\includegraphics[width=\hsize]{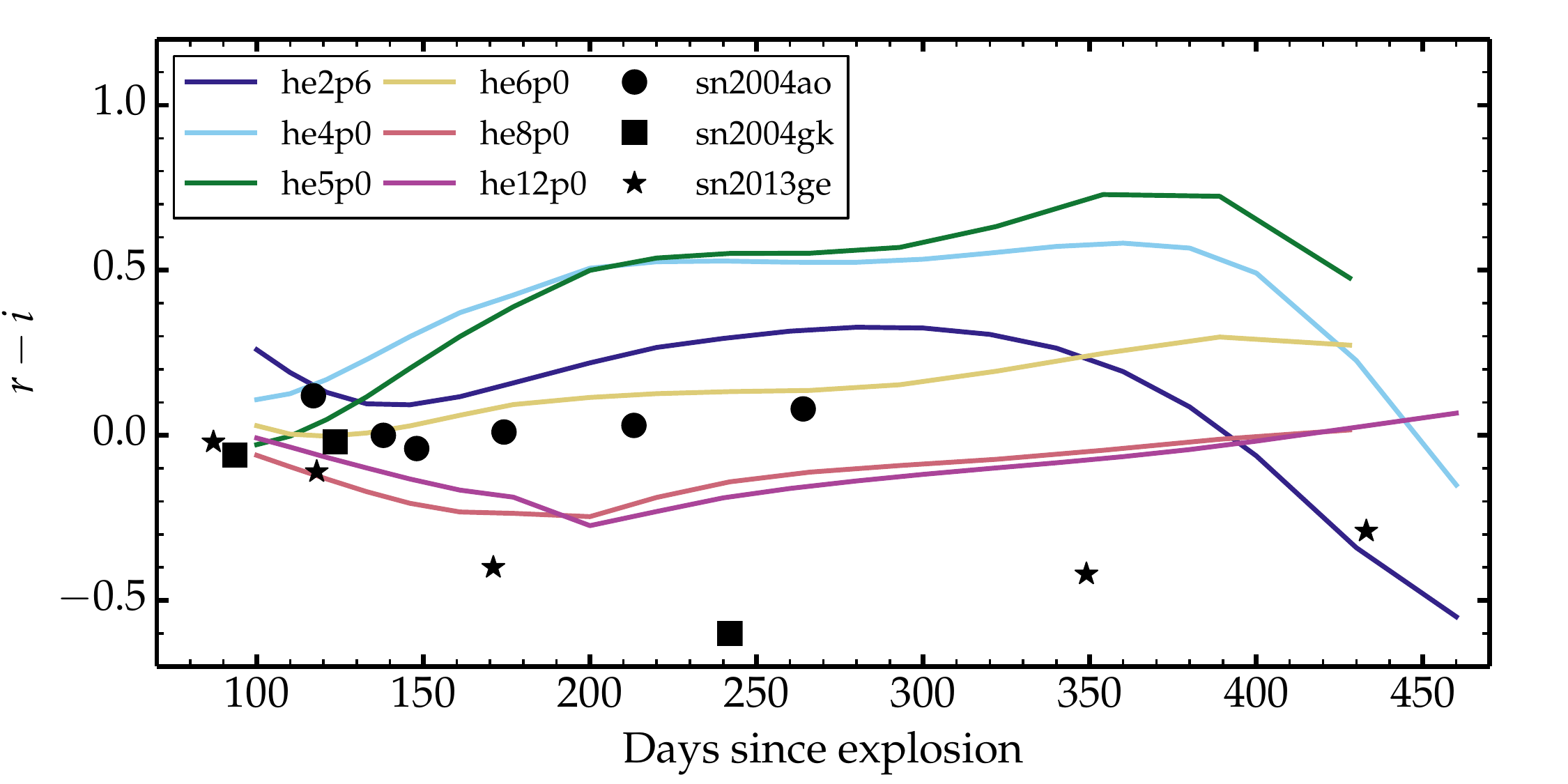}
\caption{LSST $r-i$ color evolution for the Type Ibc simulations of D21 and Dessart et al. (in preparation) together with the counterpart for Type Ib SN\,2004ao, and Type Ic SNe 2004gk and 2013ge \citep{modjaz_etal_14,shivvers_ibc_19,drout_13ge_16}. The photometry is inferred from the spectra, by convolution with the \lsst\ filter transmission curves, uncorrected for reddening or redshift -- both are small. The data were retrieved from \wiserep\ \citep{wiserep}.
\label{fig_one_color}
}
\end{figure}

Even today one struggles to obtain a well sampled spectral evolution at nebular times. Often, the spectra are truncated in the blue or in the red, so that \caiidoub\ (and especially \caiitrip) may not be observed. Many observations are stopped at around 300\,d and the coverage up to that phase is often sparse. Spectra are often noisy in distant objects, especially at later times. Figure~\ref{fig_one_color} shows the $r-i$ color evolution for a few SNe Ib and Ic, together with the small model set used here. Here, the photometry is computed from the spectra by convolution with the \lsst\ filters, which was a challenge for the reasons given above. Nonetheless, Fig.~\ref{fig_one_color} confirms that the relative strength of \oidoub\ and \caiidoub\ is captured with photometry alone.\footnote{An offset between models and observations does not imply that the method is invalid, but instead that the models are in tension with observations. The present models arise from a limited grid of theoretical models without any tuning. The numerous potential deficiencies of the models are discussed in D21, and include the adoption of He-star models initially, the assumption of spherical symmetry in both the explosion models and in the radiative transfer, the neglect of clumping and molecule formation, the large abundance of stable Nickel explosively produced etc. Figures~23--27 in D21 show that these He-star explosion models (and their variants at higher/lower explosion energies -- see D21 for details) reproduce satisfactorily the representative spectra of SNe IIb, Ib, and Ic so they evidently also reproduce the corresponding photometry.} In the future, the goal is to complement figures similar to Figs.~\ref{fig_color}--\ref{fig_one_color} with a larger set of improved progenitor and explosion models, including the SNe Ibc monitored by \lsst, as well as explore the results for various filter combinations (e.g., $u-r$, $g-r$, $r-z$ etc) since they all track different ejecta coolants.

\section{Discussion}
\label{sect_disc}

With photometry, we reduce the complexity of a nebular spectrum to just a few quantities (here mostly $g$, $r$, and $i$). Secondary lines that are often used to constrain clumping or the temperature of the various emitting regions are no longer available (e.g., \citealt{jerkstrand_15_iib}). However, we hope in the future to construct our radiative transfer models from realistic 3D explosion models that would provide a physical measure of clumping, turbulence, and mixing, rather than craft one specific model for each observed SN. Another issue is line overlap, as can occur for \oidoub\ and \niidoub\ (or even possibly H$\alpha$), and  for \caiidoub\ and [Ni\two]\,$\lambda$\,$7378$ in the 7300\,\AA\ emission feature. No information about that will be available in photometry. But even with spectra one cannot evaluate the individual contributions from complicated blends. So, the photometry will be no more ambiguous in that respect than the spectra. In contrast, photometry allows observations for a huge number of transients until late times with regular cadence and with better background subtraction than spectroscopy can provide. In that regard, photometry is far superior than spectroscopy.

As in spectroscopy, reddening will bring uncertainties. However, one may constrain the reddening from an inspection of the photometry and colors at early times, for example around maximum light in SNe Ibc \citep{drout_11_ibc,D16_SNIbc_II,woosley_ibc_21}. We show the color shift caused by a visual extinction of 1\,mag, using the Cardelli law \citep{cardelli89} in Fig.~\ref{fig_color}. While reddening shifts the models up and to the right, the distinct trajectories are retained -- it does not turn, for example, model he5p0 into model he8p0).

The observed photometric properties of a SN will be affected by its redshift. This redshift may be constrained in various ways: using existing spectroscopic redshifts of the host galaxies in the NASA/IPAC Extragalactic Database (NED) or from new host spectra; from photometric redshifts of the host galaxies (e.g., \citealt{newman_phot_redshift_22}); or from spectroscopic observations of the SN itself (e.g., \citealt{snid}). For redshift values below 0.1, most lines remain in the same filter as for zero redshift. At a redshift of 0.1, the strong lines lie between filters, in particular for \oidoub\ and \caiidoub, so the method would not then work. But for higher redshifts around 0.2, the strong lines fall again within a filter, with \oidoub\ now in $i$ and \caiidoub\ in $z$. Figure~\ref{fig_redshift} illustrates how the color-color magnitude diagram shown in Fig.~\ref{fig_color} is modified after successive redshift increments of 0.02.

At nebular times, the photometric evolution will also help in identifying some events or phenomena. Interaction with H-rich circumstellar material will lead to a strong H$\alpha$ emission that will dominate the SN radiation, which will fall primarily in the $r$ band (for zero redshift). A fast evolving light curve with a persistent and dominant brightness in $u$ and $g$  would be indicative of a Type Ibn \citep{pasto_ibn_08,hosseinzadeh_ibn_17}.

Photometric analyses, and associated statistical analyses, will play an important role in analyzing and interpreting \lsst\ SN data. However, it will still be important to obtain spectra for selected objects. For example, multiepoch spectra could greatly assist in improving photometric modeling and interpretation, and in removing biases in photometric analyses. Further, modeling of SN spectra is still under development, and thus high quality spectra (over a large passband) throughout a SN evolution, combined with high quality photometric data, will be needed to test the radiative-transfer models.

By focusing on the color evolution, one takes a more global look at the evolution and properties of SN Ibc ejecta, as opposed to a line-by-line analysis with spectra. This different, and in cases complementary approach to spectroscopy, has nonetheless numerous merits and strengths. In some sense, the debate is not whether one is superior to the other. The \lsst\ will soon be in operation and one ought to extract the information encoded in that photometric data. As illustrated here, this photometric information is very rich.

\section{Conclusion}
\label{sect_conc}

We have presented the nebular-phase radiative properties of a wide range of He-star explosions that cover both standard Type Ibc SNe and faint, currently unknown, Type Ib SNe that arise from the lowest mass He stars undergoing core collapse (\citealt{woosley_he_19}; D21).  To complement previous studies that focus almost exclusively on spectral properties (\citealt{jerkstrand_15_iib}; D21; Dessart et al., in preparation), we have focused on their photometric properties. This is motivated by the growing number of transients that have primarily photometric data, with little or no spectroscopic information. With this in mind, we framed our discussion on the \lsst\ and its $ugriz$ filters  to assess how photometry can help us constrain the progenitors and ejecta of Type Ibc SNe.

The most significant plot in this work is the $r-i$ versus $g-r$ color-color magnitude diagram for the simulations of D21 and Dessart et al. (in preparation; Fig.~\ref{fig_color}). This figure shows how each family of He-star explosions from D21 occupies a different parameter space, and how each follows a distinct trajectory over time. For redshifts below about 0.07, the filters $g$, $r$, and $i$ cover the spectral regions where the primary coolants of the various shells emit. The He-rich material, which cools through Fe\two\ emission below 5500\,\AA, is captured by the $g$ band. The O-rich material increasingly cools through \oidoub\ during the first 1--2 years after explosion, and that emission is covered by the $r$ band. Finally, radiation from the Fe-rich and Si-rich material comes out primarily in \caiidoub, which falls within the $i$ band. Different proportions of these various shell masses lead to a distinct color evolution. What is lost with the lack of spectra is also compensated by the much greater monitoring at all epochs with the \lsst. The $ugrizy$ filter set is also much more extended in wavelength than most SN spectra, which are typically truncated in the blue and often in the red.

The photometric approach will also allow SNe to be studied until much later times, an endeavor that has been possible only with the most nearby events such as the Type IIb SN 1993J. Further, it will be possible  to capture the onset of an interaction with circumstellar material, as evidenced through a sudden reduction of $u-r$ color \citep{DH_interaction_22}.

Obviously, degeneracies that affect SN Ibc spectra will also affect their photometry. Recently, \citet{fang_neb_22} and \citet{prentice_neb_22} presented an analysis of nebular-phase spectra of SNe Ibc and revealed the absence of a clear correlation between the flux ratio of \oidoub\ and \caiidoub\ in their Type Ibc SN sample, which conflicts with the common expectation that Type Ib and Type Ic SN progenitors have different (increasing) O content. Obviously, a photometric analysis would similarly reveal a lack of correlation in $r-i$ for this sample. Physically, there may be various reasons for this lack of correlation. First, while SNe Ic may arise from higher mass progenitors on the main sequence than SNe Ib, the effect of wind mass loss might lead to a relatively narrow range of preSN masses (D21). Secondly, the sample of observed SNe Ibc seems to lack low-mass He-rich ejecta with very weak \oidoub\ (corresponding to our models he2p6--he4p0) while higher mass ejecta with several solar masses of O are unfavored by the initial mass function and may thus be more scarce. What may then drive the currently observed diversity between and amongst SNe Ib and Ic may not be a composition difference, as generally assumed, but instead the large scale macroscopic mixing of \nifs\ and other elements like He and O \citep{d12_snibc}.  Further work is needed to address this issue in stripped-envelope SNe. Crucially, \lsst\ photometry, and color-color diagrams formed from that photometry, will provide strong statistical constraints on SN properties and the progenitors that give rise to these SNe, and hence can directly address the issue.

\begin{acknowledgements}

This work was supported by the ``Programme National de Physique Stellaire'' of CNRS/INSU co-funded by CEA and CNES. Support for JLP is provided in part by ANID through the Fondecyt regular grant 1191038 and through the Millennium Science Initiative grant  CN12\_009, awarded to The Millennium Institute of Astrophysics, MAS.  DJH thanks NASA for partial support through the astrophysical theory grant 80NSSC20K0524. H.K. was funded by the Academy of Finland projects 324504 and 328898. EDH acknowledges support from ANID PhD scholarship No. 21222163. This work was granted access to the HPC resources of  CINES under the allocation 2020 -- A0090410554 and of TGCC under the allocation 2021 -- A0110410554 made by GENCI, France. This research has made use of NASA's Astrophysics Data System Bibliographic Services.

\end{acknowledgements}


\appendix

\section{Additional figures}

\begin{figure}
\centering
\includegraphics[width=\hsize]{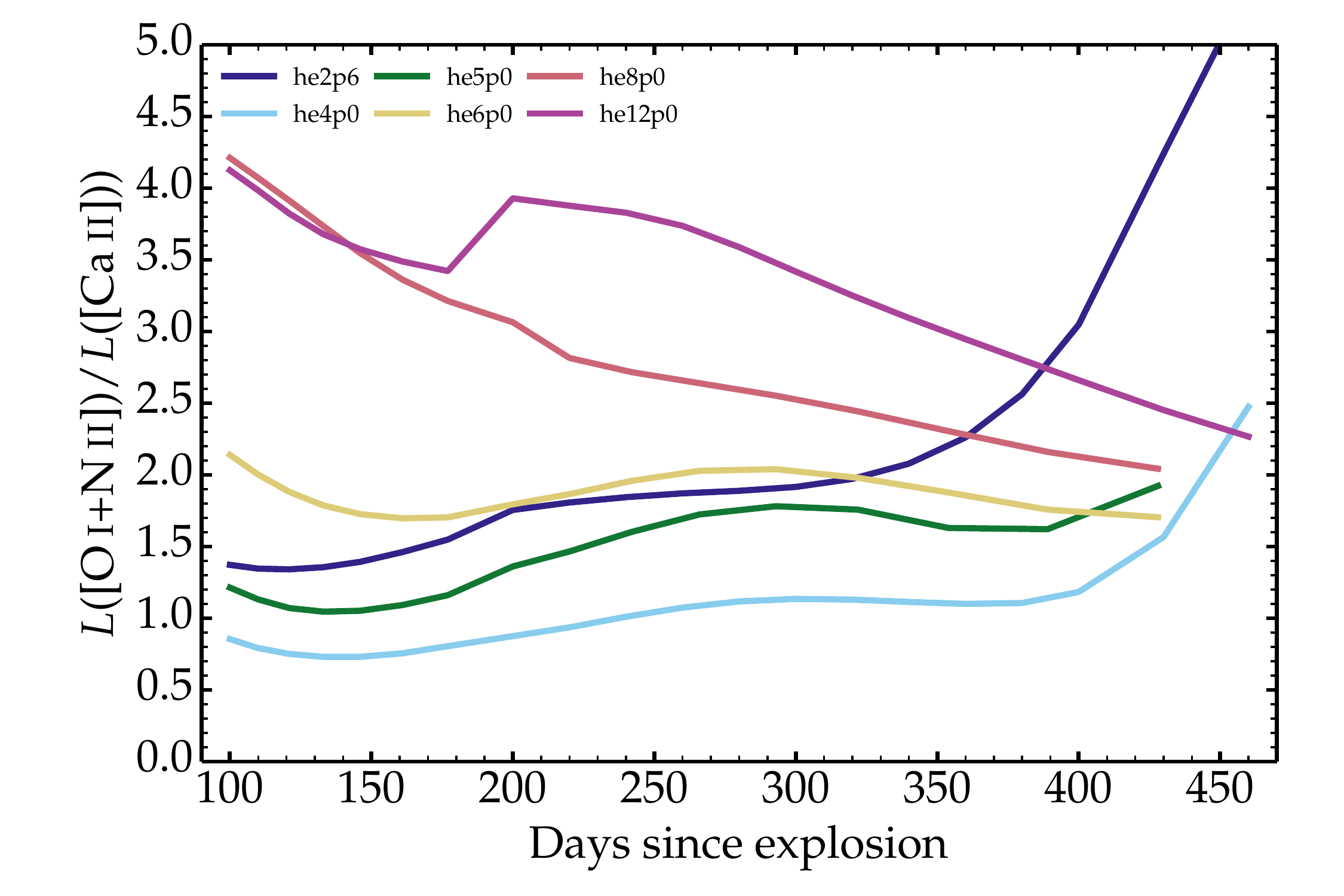}
\caption{Evolution of the ratio of the combined \oidoub\ and \niidoub\ fluxes with that of \caiidoub\ for models he2p6, he4p0, he5p0, he6p0, he8p0, and he12p0. These curves are analogous to the $r-i$ color curves shown in Fig.~\ref{fig_phot} and Fig.~\ref{fig_one_color}.
\label{fig_oi_nii_over_caii}
}
\end{figure}

\begin{figure*}
\centering
\includegraphics[width=0.49\hsize]{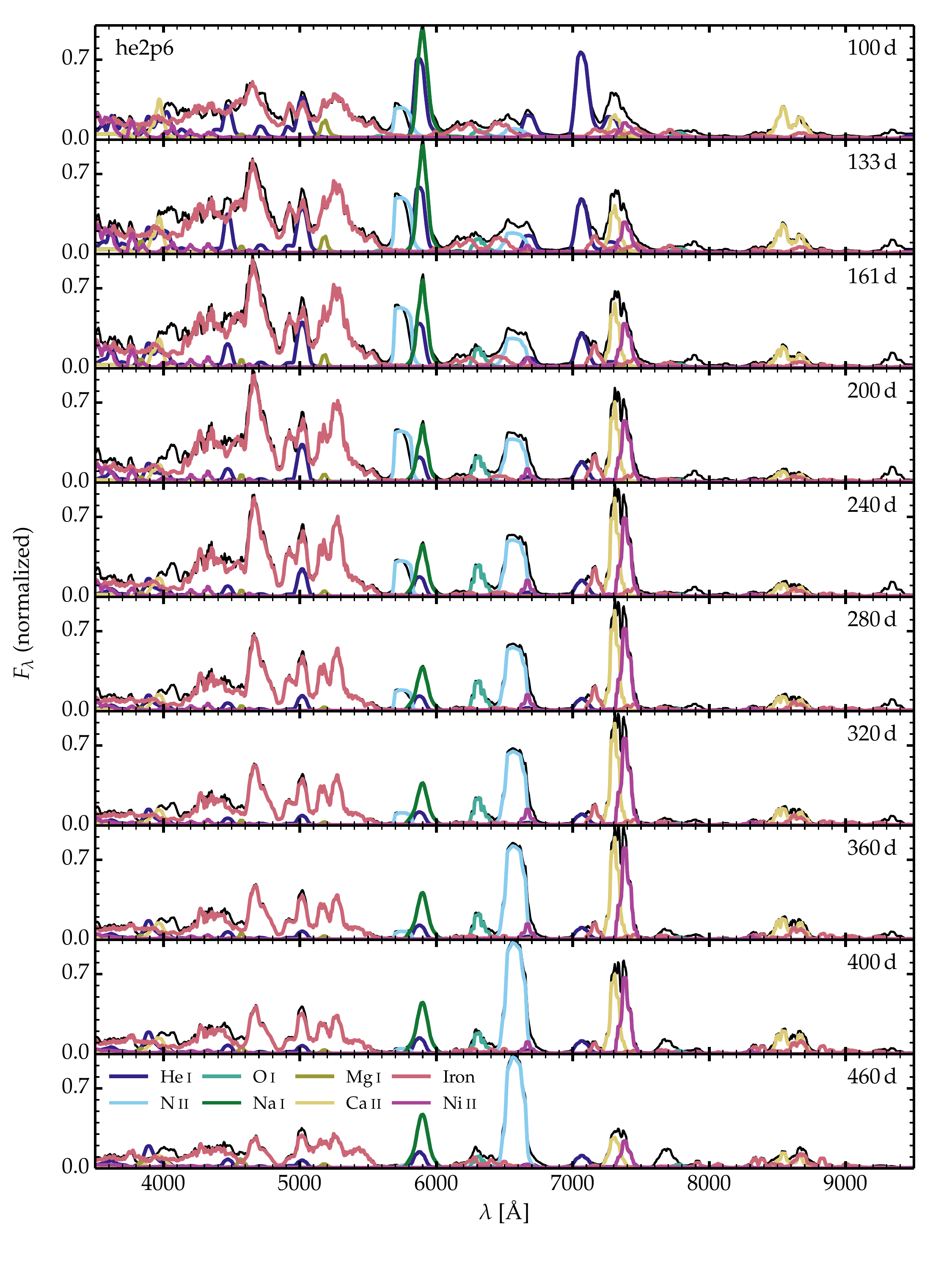}
\includegraphics[width=0.49\hsize]{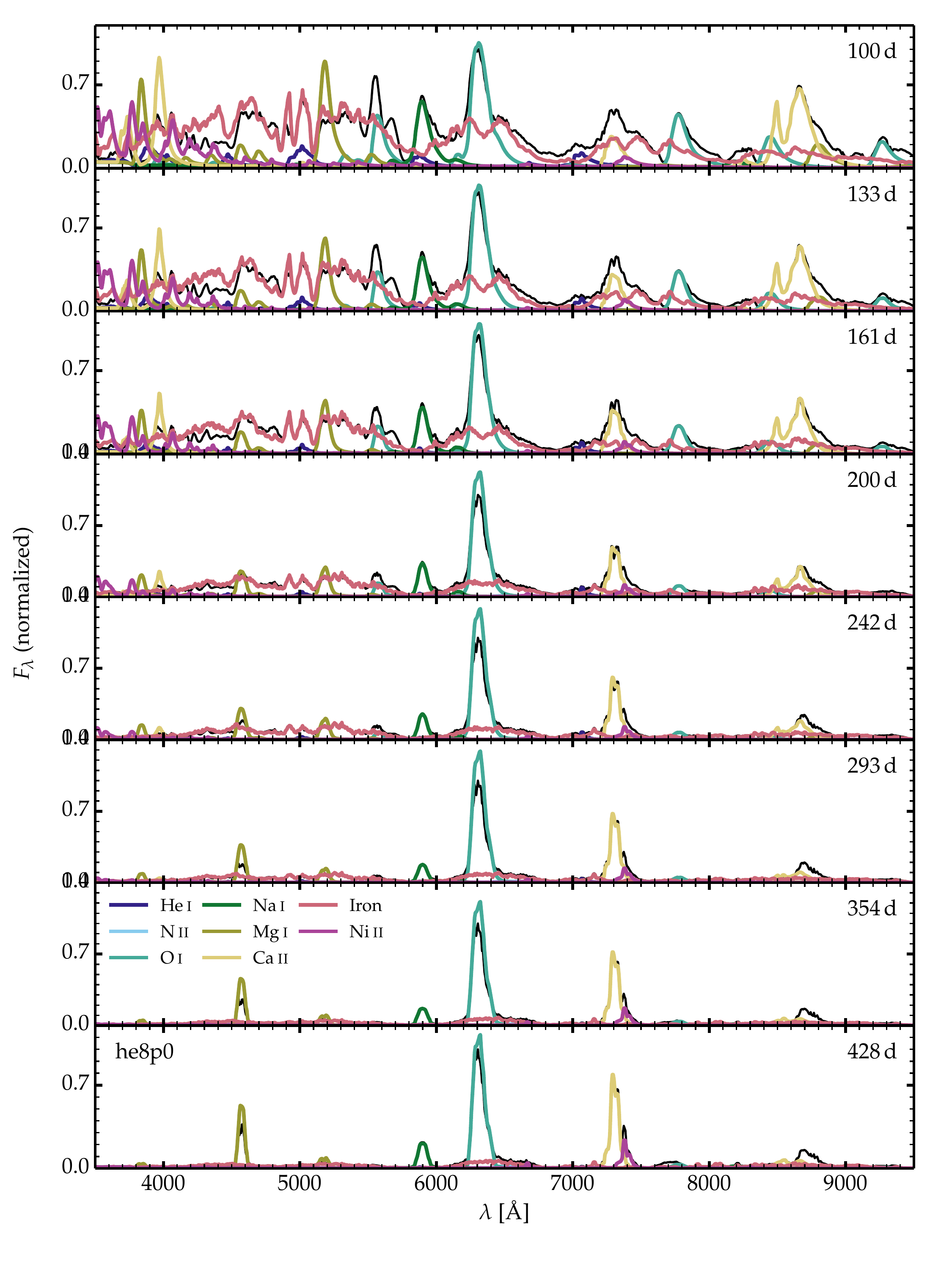}
\caption{Evolution of the optical spectrum of model he2p6 (left) and he8p0 (right) from 100 to about 450\,d. Overplotted is the contribution from bound-bound transitions associated with various ions. The spectrum that includes all Fe\one, Fe\two, and Fe\three\ bound-bound transitions is labeled by the key ``Iron".
\label{fig_spec}
}
\end{figure*}

Figure~\ref{fig_oi_nii_over_caii} illustrates the close correspondence between the ratio of the combined \oidoub\ and \niidoub\ fluxes with that of \caiidoub\ and the color $r-i$ shown in Fig.~\ref{fig_phot}. If we were to omit the lower mass He-star explosion models, the color $r-i$ would reflect the ratio of the \oidoub\  and the \caiidoub\ line fluxes, with a slight offset due to the neglect of the overlapping iron and nickel emission (D21; Fig.~\ref{fig_spec}).

Figure~\ref{fig_spec} shows the spectral evolution for the low-mass model he2p6 and the high-mass model he8p0 from 100 to about 450\,d after explosion. A detailed description of the spectral properties is given in D21 and in Dessart et al. (in preparation).

\begin{figure*}
\centering
\includegraphics[width=0.9\hsize]{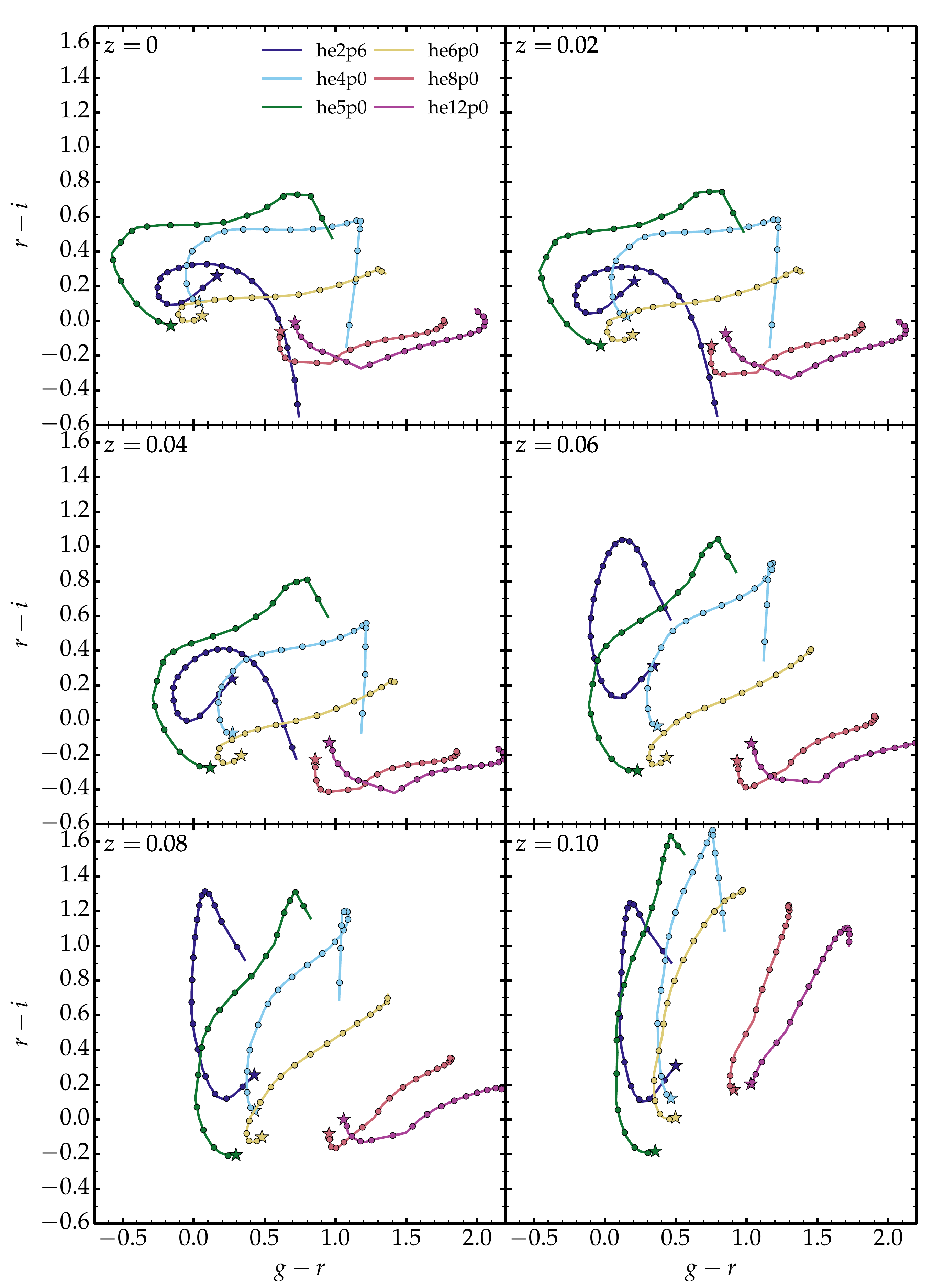}
\caption{Same as Fig.\ref{fig_color} but for redshift increments of 0.02
from 0 at top left to 0.1 (bottom right).
\label{fig_redshift}
}
\end{figure*}

 Figure~\ref{fig_redshift} shows how the model trajectories in the color-color magnitude diagram of Fig.~\ref{fig_color} are modified after successive redshift increments of 0.02. For redshifts below about 0.06, the strongest spectral lines remain within the same \lsst\ filter as for a redshift of zero. Increasing the redshift to 0.1 leads to significant changes because the strong emission lines in $r$ and $i$ approach and cross the edge of these filters. Nonetheless, the respective models are still distinguishable.

\end{document}